# A recoverable versatile photo-polymerization initiator catalyst

**Dianyu Chen,**[a] **Rongxin Yuan**[a] **and Soumyajit Roy*[a,b]**

A photo-polymerization initiator based on an imidazolium and an oxometalate, viz., $(BMIm)_2(DMIm)\ PW_{12}O_{40}$ (where, BMIm = 1-butyl-3-methylimizodium, DMIm = 3,3'-Dimethyl-1,1'-Diimidazolium) is reported. It polymerizes several industrially important monomers and is recoverable hence can be reused. The $M_n$ and PDI are controlled and a reaction pathway is proposed.

Fast receding reserve of resources and growing demand of 'green' plastics need an immediate synthetic technique that can provide plastics but is low on reserve of resources. Although photo-polymerization is a step forward in that direction, however the condition of repeated addition of initiator often limits its viability for large scale production. Development of a versatile yet recoverable photo-polymerization initiator catalyst would be a welcome step forward in that direction. Here we report such an initiator catalyst. Logically such an initiator catalyst would contain a photoactive radical generating component, a radical propagating component, an oligomer stabilizing hydrophobic component and an initiator catalyst regenerating component.

Well documented photo-catalytic activity of $PW_{12}O_{40}^{3-}$ type oxometalates (OM) prompted its choice as a photoactive radical generating component of the initiator catalyst.[1-5] Certain interesting structural and electronic features of BMIm type ionic liquids led to the choice of $BMIm^+/DMIm^+$ as another component of the initiator catalyst.[6-8] For instance: 1. Ability to provide the reaction system with a radical propagating source (imidazolium moiety). 2. As in ionic liquids, they can form extended hydrophobic channel like compartments to stabilize emerging oligomers.[7,8] Moreover the OM's ability of redox regeneration is also exploited. In short, our design strategy of the initiator catalyst exploits 1. the photoactivity of oxometalate to initiate polymerization; 2. combines it with all the advantages of the imidazolium moiety needed to render the initiator catalyst functional and 3. redox recoverability of OM under the stipulated reaction condition.

We now look how the polymerization reaction takes place.[9] (‡ For synthetic details of the initiator catalyst refer to ESI.) 20 mg of the initiator catalyst, $(BMIm)_2(DMIm)\ PW_{12}O_{40}$ is mixed with ca. 3 – 5 ml of the monomer (styrene, vinyl isobutyl ether, α-methyl styrene, 1,1-dicyanoethylene, methyl methacrylate, methyl acrylate, butyl methacrylate, butyl acrylate, vinyl acetate, maleic anhydride). The mixture is dissolved in 20 ml acetone. The colourless mixture is then put inside a photo-reactor tube and of a XPA-1 photo-reactor, operating at 300 W and λ=365 nm for 30 minutes. After 30 minutes, the contents of the tube become faint blue in colour. (See ESI.) The tube is taken out and the contents are transferred to a round-bottom flask. 200 ml of dichloromethane is added to the contents and the solvent is slowly removed. The polymer, extracted in the layer with dichloromethane, is separated and characterized by GPC, while the initiator catalyst remains dispersed and un-dissolved. The initiator catalyst containing fraction upon drying undergoes several changes of colour (like, from blue to yellow to dirty yellow) till a colourless white powder is formed, upon exposure to air. (See ESI.) The initiator catalyst is recovered. The recovery of the initiator catalyst was tested tediously gravimetrically by weighing the recovered initiator catalyst and checking the FTIR spectra of the recovered material. This recovery ranges from 40% to 50% of the starting amount. (Fig. 1) Interestingly the loss of the initiator catalyst is highest in the first cycle for all monomers and thereafter it reaches almost a constant value till the tenth cycle. (Fig. 1)

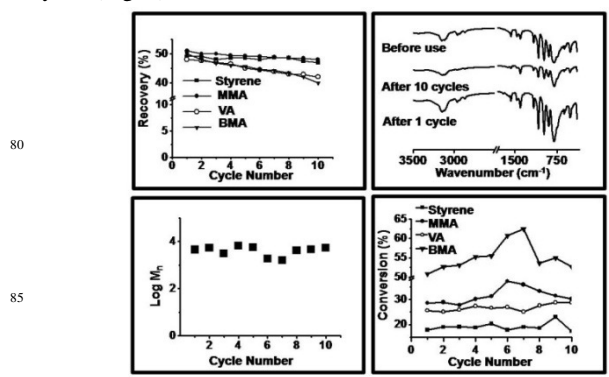

**Fig. 1** Recoverability (top-left), FTIR spectra of the recovered initiator catalysts (top-right), $M_n$ of the polymers obtained over several cycles (bottom-left) and corresponding conversions (bottom-right).

An analysis of initiator catalyst recovery from the stand point of molecular integrity using FTIR spectroscopy is described later.

[a] *School of Chemistry and Materials Engineering, Changshu Institute of Technology, Changshu, Jiangsu, P. R. China.*
*Fax: +8651252251821*
[b] *Eco-friendly Applied Materials Laboratory, DCS, New Campus, Indian Institute of Science Education and Research, Kolkata, India.*
*Fax: +913325873020, E-mail: roy.soumyajit@googlemail.com, s.roy@iiserkol.ac.in*
†Electronic Supplementary Information Available: Synthesis and characterization of the initiator catalyst, DFT calculation details and plots.

Note: $^1$H NMR spectroscopy did not lead to the analyses of the recovery of initiator catalysts probably due to the generation of paramagnetic centres. Also note, the recovered initiator catalyst is further used in successive polymerization cycles and it acts without any significant loss of polymerization power and without the initiator catalyst no polymerization takes place. (Fig. 1) The $M_n$ of polymers so obtained, the corresponding conversions over successive cycles of polymerization using the same initiator catalyst are shown and the results reflect a high degree of recoverability of the initiator catalyst. (Fig. 1)

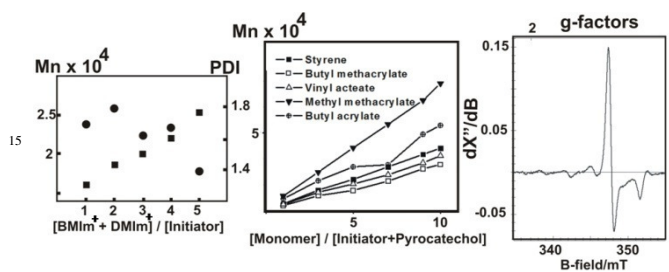

**Fig. 2** Left: Increasing $M_n$ (■) and decreasing PDI (●) with increasing BMIm$^+$ and DMIm$^+$ concentration to a fixed initiator catalyst concentration during butyl methacrylate polymerization. (Same trend is obtained with other monomers.) Middle: Increasing $M_n$ with increasing pyrocatechol loading for different monomers. (Same result is obtained with other dihydroxy phenols.) Right: EPR spectra of a reaction mixture initiated by (BMIm)$_2$(DMIm) PW$_{12}$O$_{40}$ after 10 minutes of reaction in acetone.

We now investigate the effect of the initiator catalyst concentration on the $M_n$ and PDI of the synthesized polymers, to understand the underlying pathway of this polymerization reaction. With this end in view polymerization reactions were carried out increasing the relative concentration of an equimolar mixture of BMIm$^+$ and DMIm$^+$ with respect to the initiator catalyst. (Note in these cases chloride was used as counter-ion and the effect does not change with other halides.) It is observed that with increasing concentration of BMIm$^+$ and DMIm$^+$ the $M_n$ of the synthesized polymers increases whereas the PDI decreases. (Fig. 2) This implies that with increasing BMIm$^+$ and DMIm$^+$ concentration the relative concentration of the radicals responsible for polymerization decreases. In other words, this observation indicates that a free-radical polymerization takes place and the radical source is directly related to the oxometalate part of the initiator catalyst. To prove the participation of OM• radicals we added to the reaction mixture o-, m- and p- dihydroxy phenols in increasing ratio with respect to the initiator catalyst. It is observed that with an increase in loading of the dihydroxy phenols, (o-, m- or p-) the $M_n$ of the polymers increase. (Fig. 2) This is so because with an increase in dihydroxy phenol loading more of oxometalate radicals are scavenged, a fact well known in the literature.[1-5] Consequently, lesser radicals are available for polymerization leading to high $M_n$ polymers. This observation in turn also shows clearly the participation of oxometalate radicals in polymerization. (Fig. 2) Time resolved EPR spectroscopy finally conclusively proves this conjecture of oxometalate free-radical participation in initiating the polymerization reaction by showing the existence of oxometalate radical in polymerization initiation. The generation of a paramagnetic OM• radical (S = 1/2) upon irradiation is demonstrated by the EPR spectrum. (Fig. 2) The second peak at g = 1.93 perhaps corresponds to imidazolium cation radical bound to OM. It hence seems that any exposure of OM(ox) to UV light result in generation of OM in excited state (OM*) and OM• radical irrespective of the presence of other materials in the reaction mixture, which in turn leads to the generation of an imidazolium radical cation.

Putting the pieces of the evidences together an overall picture of the polymerization pathway emerges. Initially upon irradiation, photo-activation of the inorganic part of the initiator catalyst leads to the formation of OM* and eventually an OM• radical, which in turn generates an imidazolium radical cation which can effectively initiate free-radical/cationic photo-polymerization, rendering the initiator catalyst its versatility and OM in turn itself gets reduced to form HOM.[10] Generation of such a radical cation hinges on the transfer of hydrogen free radical to irradiated OM in excited state forming reduced HOM which on exposure to air regenerates the active OM. The successive change of colour of the used initiator catalyst during drying and recovery implies such step-wise oxidation of the HOM to OM.

DFT calculations furthermore reveal band-gap energy of 2.729 eV, and justify photo-excitation of the anionic oxometalate part of the initiator catalyst to form a radical upon irradiation. (See ESI.) DFT furthermore shows that the OM cluster surface has high degree of electrophilicity and the electrophilic $C_3$ centres of the cluster surface also play an important role in polymerization reaction by acting as a docking site for the transfer of H• of imidazolium moiety to OM during polymerization cycles.

Still open questions remain: how do the emerging hydrophobic polymers stabilize themselves? Does the initiator catalyst also have a stabilizing role? The single crystal X-ray diffraction pattern of the initiator catalyst reveals continuous elliptical hydrophobic channels 7 Å in diameter, which are in close vicinity to the PW$_{12}$O$_{40}$$^{3-}$ – OM core. (Fig. 3) We speculate that these channels act as stabilizing hydrophobic compartments for the growing polymer chains and thereby facilitate formation of hydrophobic polymers with long chain length. It is to be noted that alone PW$_{12}$O$_{40}$ with only protons/ammonium as counter-ions although can act as an initiator catalyst however leads to polymers with extremely high PDI and very low $M_n$ and are not recoverable. This observation supports our proposition that butyl channels of imidazolium provides a stabilizing hydrophobic channel and thereby enables successful propagation by the initiator catalyst.

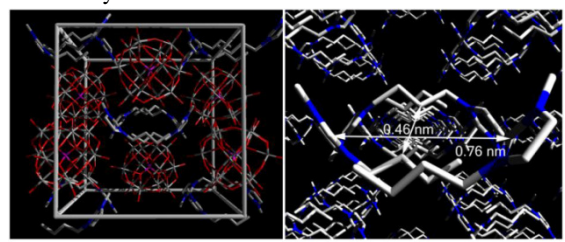

**Fig. 3** Unit cell (viewed along c-axis) of the initiator catalyst (left); and the continuous elliptical hydrophobic channel (right) (Nitrogen, blue; carbon, light-grey; oxygen, red; phosphorus, lilac; tungsten, grey.)

We finally wanted to investigate the exact extent of the recovery of the initiator catalyst. We separated the recovered initiator catalyst and weighed it and checked the FTIR spectra. From the FTIR spectra we wanted to further quantify the extent of recovery of the initiator catalyst on a molecular level. To do so, we analysed the finger-print regime of the initiator catalyst. It is known that [α-$PW_{12}O_{40}$] type oxometalates have following characteristic vibrational frequencies: 804 and 894 cm$^{-1}$ ($v_{as}$,W-O-W), 979 cm$^{-1}$ ($v_{as}$,W=O(t)), 1078 cm$^{-1}$ ($v_{as}$,P-O(br)).[5] Likewise, peaks of BMIm$^+$ and DMIm$^+$ at 2867, 2958, 3116 and 3147 cm$^{-1}$ are assigned to ($v_s$,CH$_3$), ($v_{as}$,CH$_3$), [$v_s$,HC(4)-C(5)H] and [$v_{as}$,HC(4)-C(5)H] respectively.[7] It is reasonable to believe that retention of the finger-print implies retention of the molecular structure of the initiator catalyst. In addition, we wanted to investigate the exact extent of retention of the original molecular structure of the initiator catalyst by finding the intensity ratio of the corresponding frequencies. One such recovery detail is shown schematically. (Fig. 4) We observe that there is a component level discrepancy in recovery of the initiator catalyst. Although 40%-80% of the oxometalate component indeed retains the molecular structure, only ca. 10-25% of the imidazolium component retain the molecular structure at the end of 10 cycles as reflected by the reduction in related intensity ratios. Hence it appears that a significant part of DMIm$^+$ and BMIm$^+$ from the initiator catalyst lattice is depleted in course of the polymerization reaction cycles. This loss after first cycle is slow and asymptotic in nature and allows polymerization reaction to continue over more number of cycles. A reasonable correlation was further found between the molecular level recoverability (as ascertained above from the ratios of various characteristic FTIR frequencies) and gravimmetric recoverability (by weight and checking the FTIR finger-print) of the initiator catalyst which ranges between 40%-50%.

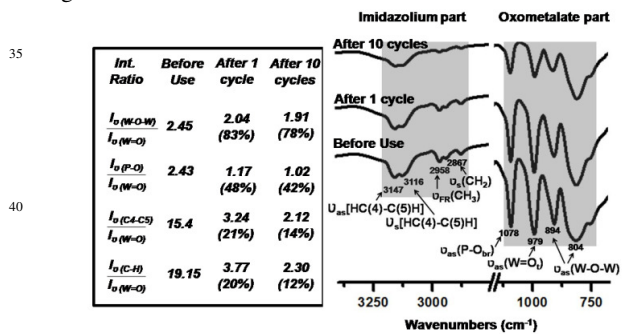

**Fig. 4** FTIR spectroscopic investigation on the recovery and loss of the initiator catalyst. Oxometalate and imidazoilum regimes shown.

To summarize, upon UV-irradiation the initiator catalyst (BMIm)$_2$(DMIm) PW$_{12}$O$_{40}$ generates a OM in excited state (OM*). Reason: the HOMO-LUMO gap corresponds to the wavelength of irradiation. An H• is then transferred from imidazolium cation to OM in excited state to generate an imidazolium radical cation and thereby reducing OM* to HOM. Reason: vicinity of the imidazolium hydrogen to OM* center and acidity of the corresponding hydrogen. This reduction of the OM* centre also manifests in the change of colour of the initiator catalyst from colourless to blue. The generated imidazolium radical cation then initiates further free-radical/cationic polymerization, rendering the initiator catalyst its versatility. In the following steps the lost imidazolium cation is replenished by the lattice DMIm$^+$ and BMIm$^+$. This is reflected in the depletion of the total imidazolium content from FTIR spectroscopic experiments. The reduced HOM on the other hand is stepwise oxidized and regenerated in course of solvent removal/drying as evidenced by successive colour change to the starting oxidized OM. The complete reaction pathway is summarized in Fig. 5.

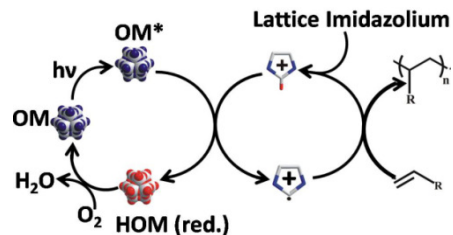

**Fig. 5** Overall scheme of the polymerization reaction: Generation of OM* (OM inexcited state) is shown which in turn generates imidazolium cation radical by abstracting hydrogen and itself gets reduced to HOM. The HOM is further oxidized by aerial oxygen to give the starting OM.

This reaction can run efficiently over many cycles and thereby polymerizing an array of monomers, like, styrene, vinyl isobutyl ether, α-methyl styrene, 1,1-dicyanoethylene, methyl methacrylate, methyl acrylate, butyl methacrylate, butyl acrylate, vinyl acetate, maleic anhydride. Of these monomers, maleic anhydride by this route undergoes homopolymerization, an observation also interesting in itself.

The authors thank Dr. W. Guan and Professor Z. M. Su of Northeast Normal University, Changchun, Jilin, P. R. China, for their help with the DFT calculations; and IISER-K, India, and CIT, P. R. China for financial support.

## Notes and references

‡ *Synthesis of the initiator catalyst, its gravimmetric recovery method and other crystallographic§ and spectroscopic details along with details of DFT calculation are included in the ESI.*

1  E. Androulaki, A. Hiskia, D. Dimotikali, C. Minero, P. Calza, E. Pelizzetti, E. Papaconstantinou, *Environ. Sci. Technol.* 2000, **34**, 2024 – 2028.
2  H. Hori, E. Hayakawa, H. Einaga, S. Kutsuna, H. Kiatagawa, S. Arakawa, *Environ. Sci. Technol.* 2004, **38**, 6118 – 6124.
3  A. Hiskia, M. Ecke, A. Troupis, A. Kokorakis, H. Hennig, E. Papaconstantinou, *Environ. Sci. Technol.* 2001, **35**, 2358 – 2364.
4  A. Hiskia, E. Androulaki, A. Mylonas, A. Troupis, E. Papaconstantinou, in *Polyoxometalate Chemistry*, (Eds: M. T. Pope, A. Müller) Kluwer Academic Publishers, 2001, pp. 417 – 424.
5  M. T. Pope, *Heteropoly and isopoly oxometalates*; Springer: Berlin, 1983; M. T. Pope, A. Müller, *Polyoxometalate chemistry: From topology Via self-assembly to applications*, Kluwer Academic Publishers: Dordrecht, London, 2001; C. L. Hill (Ed.) *Chem. Rev.* 1998, **98 (1)**.
6  T. Welton, *Chem. Rev.* 1999, **99**, 2071-2084.
7  P. Wasserscheid, T. Welton, Eds. *Ionic Liquids in Synthesis*; Wiley-VCH: Weinheim, Germany, 2002.
8  H. Ohno, Ed. *Electrochemical aspects of ionic liquids*; Wiley: New York, 2005.
9  Patent application filed to protect the invention: Application Number: CN 201010297714.5.
10 S. Kim, H. Park, W. Choi, *J. Phys. Chem. B.* 2004, **108**, 6402 – 6411.